\documentclass{article}

\usepackage[utf8]{inputenc}

\usepackage[english]{babel}

\usepackage{epsfig,graphicx,graphics}
\usepackage{color}
\usepackage{mathtools}
\usepackage{amssymb, amsmath}

\usepackage[percent]{overpic}

\graphicspath{ {./} }

\newtheorem{definition}{Definition}
\newtheorem{theorem}{Theorem}
\newtheorem{lemma}{Lemma}

\newtheorem{corollary}{Corollary}
\newtheorem{remark}{Remark}
\newtheorem{conjecture}{Conjecture}

\usepackage{subcaption}
\usepackage{float}
\usepackage{comment}

\title{Inversion and Symmetries of the Star Transform}
\author{G. Ambartsoumian$^\dagger$ {\small and} M. J. Latifi Jebelli$^*$\\
\small $^\dagger$Department of Mathematics, University of Texas at Arlington\\
\small $^*$Department of Mathematics, University of Arizona}

\date{}

\begin{document}

\maketitle

\begin{abstract}
The star transform is a generalized Radon transform mapping a function of two variables to its integrals along ``star-shaped'' trajectories, which consist of a finite number of rays emanating from a common vertex. Such operators appear in mathematical models of various imaging modalities based on scattering of elementary particles. The paper presents a comprehensive study of the inversion of the star transform. We describe the necessary and sufficient conditions for invertibility of the star transform, introduce a new inversion formula and discuss its stability properties. As an unexpected bonus of our approach, we prove a conjecture from algebraic geometry about the zero sets of elementary symmetric polynomials.
\end{abstract}

\section{Introduction}

The star transform is a generalized Radon transform mapping a function in $\mathbb{R}^2$ to its weighted integrals along  ``star-shaped'' trajectories, which consist of a finite number of rays emanating from a common vertex. Such operators appear in mathematical models of single-scattering optical tomography and single-scattering X-ray CT \cite{Florescu2, Florescu3, Florescu4, Kats-Kryl-15, ZSM-star-14}. Particular attention has been drawn to the transforms with only two rays, usually called either broken ray transforms or V-line transforms \cite{Amb1, Amb-Latifi1, Amb-Moon, Amb-Roy, Florescu1, Florescu4, Gouia-Amb, Kats-Kryl-13, Kats-Kryl-15, Sherson-15, Terz, Walker_2019}. In addition to that, multiple authors have studied various generalizations of these transforms to higher dimensions (see the review article \cite{Amb2} and the references there).

All works cited above deal with cases, where the vertices of the stars or V-lines are inside the image domain. This paper also studies the star transform in that setup. A whole other world of applications and mathematical results exists in the case, when the vertices of the V-lines (or cones in higher dimensions) are restricted to the boundary of the image domain (e.g. see \cite{Haltm-Moon-Schief, Voichita, Ngu-Tru-Gran, TruongNguyen1, TruongNguyen2}). To find out more about the latter, we refer the interested reader to the review articles \cite{Amb2, TerzKuchKun} and the references there.

While there are multiple interesting formulas and procedures for inversion of the V-line transform, the star transform with three or more branches has been studied only in the pioneering paper \cite{ZSM-star-14}. The method presented in that article uses Fourier analysis techniques to reduce the inversion of the star transform to a problem of solving an infinite system of linear equations. The latter is solved using various truncations of the system and fast iterative computations of the Tikhonov-regularized pseudo-inverses of the resulting finite-size rectangular matrices. The paper also discusses the stability of the proposed inversion algorithms, formulating an algebraic condition that is necessary and sufficient for the stable reconstruction. Finally, several special geometric setups of the star transform are studied in relation to that condition.

This paper presents a comprehensive study of the inversion of the star transform using a totally different approach. We introduce a new exact closed-form inversion formula for the star transform, by establishing its connection with the (ordinary) Radon transform. It is much simpler than the inversion algorithm described in \cite{ZSM-star-14} and provides a very intuitive geometric insight to the problem. Our formula leads to a description of necessary and sufficient conditions for invertibilty of the star transform based on its geometric configuration, which was not known before. It also naturally yields the necessary and sufficient condition for the stability of inversion described in \cite{ZSM-star-14}. A detailed analysis of this condition allows us to classify the geometric setups of the transform according to their stability and provides a recipe for building star configurations with the ``most stable'' inversions. As an unexpected bonus of our analysis, we prove a long-standing conjecture from algebraic geometry (see \cite{Conflitti}) about the zero sets of elementary symmetric polynomials.

The rest of the article is organized as follows.
In Section \ref{sec:def} we define the main concepts used in the paper. In Section \ref{sec:inversion} we present a new  inversion formula for the star transform. In Section \ref{sec:inj} we study the injectivity of that transform. In Section \ref{sec:sing-dir} we discuss the so-called singular directions of the star transform that play an important role in the stability of numerical reconstructions. In Section \ref{sec:conj} we state and prove the aforementioned conjecture from algebraic geometry. In Section \ref{sec:applications} we describe mathematical models of some imaging modalities that use the star transform. In Section \ref{sec:numerical} we present various numerical simulations demonstrating the efficiency of our inversion method. We finish the paper with some additional remarks in Section \ref{sec:remarks} and acknowledgements in Section \ref{sec:acknowledge}.

\section{Definitions}\label{sec:def}

\noindent
Let $f(x)\in C_c\left(\mathbb{R}^2\right)$ be a compactly supported continuous function, and let $\gamma$ be a unit vector in the plane.

\begin{definition}
 The \textbf{divergent beam transform} $\mathcal{X_\gamma}$ of $f$ at $x\in \mathbb{R}^2$ is defined as:
 \begin{equation}\label{def:DivBeam}
   \mathcal{X}_{\gamma}f(x) =  \int_{0}^{\infty} f(x+t \gamma)\,dt.
 \end{equation}
 \end{definition}

One can use a linear combination of divergent beam transforms with a set of fixed distinct directions $\gamma_i\in \mathbb{S}^1$ and non-zero constants $c_i\in\mathbb{R},\, i=1,\ldots,m$ to define the star transform of $f$. Namely,

\begin{definition}
 The (weighted) \textbf{star transform} $\mathcal{S}$ of $f$ at $x\in \mathbb{R}^2$ is defined as:
 \begin{equation}\label{def:Star}
  \mathcal{S}f(x) = \sum_{i=1}^{m}\; {c_i\;\mathcal{X}_{\gamma_i}f(x)} = \sum_{i=1}^{m}\; {c_i\int_{0}^{\infty} {f(x+t \gamma_i)\,dt}}.
 \end{equation}
 \end{definition}

\begin{figure}[ht]
\begin{subfigure}{.5\textwidth}
  \centering
 \includegraphics[height=1.5in]{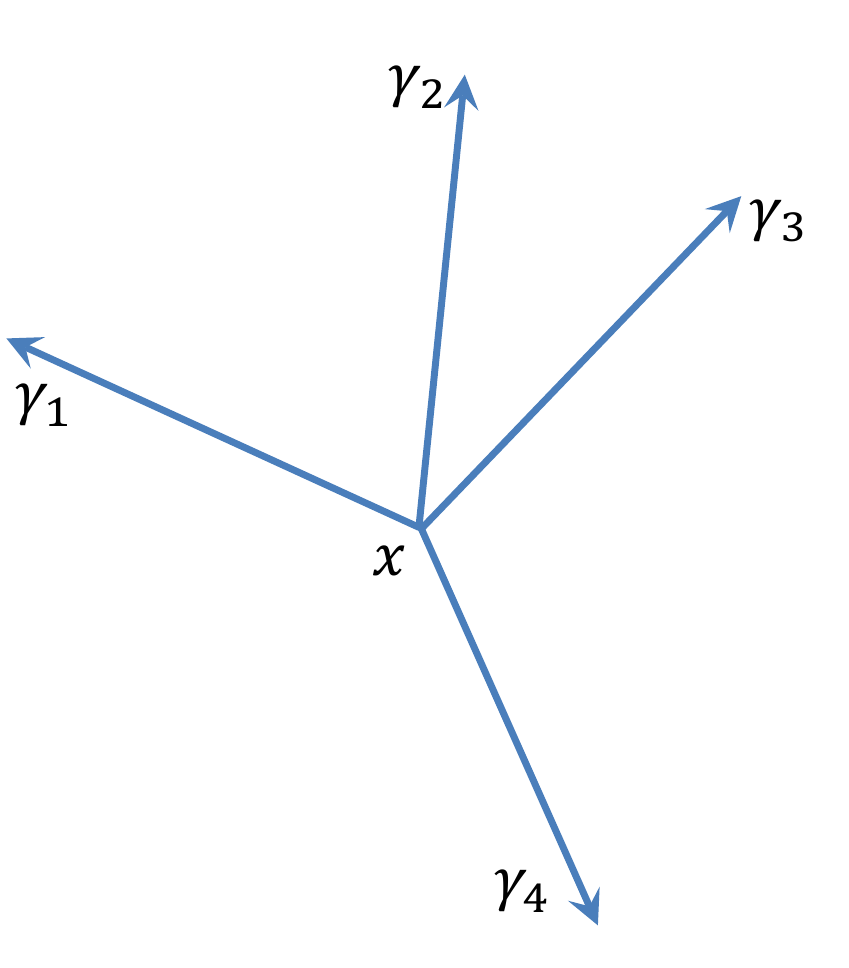}
 \caption{ A star centered at point $x$\\ with rays emanating along fixed\\ unit vectors $\gamma_1,\ldots, \gamma_4$. }
  \label{fig1a}
\end{subfigure}
\begin{subfigure}{.5\textwidth}
  \centering
\includegraphics[height=1.5in]{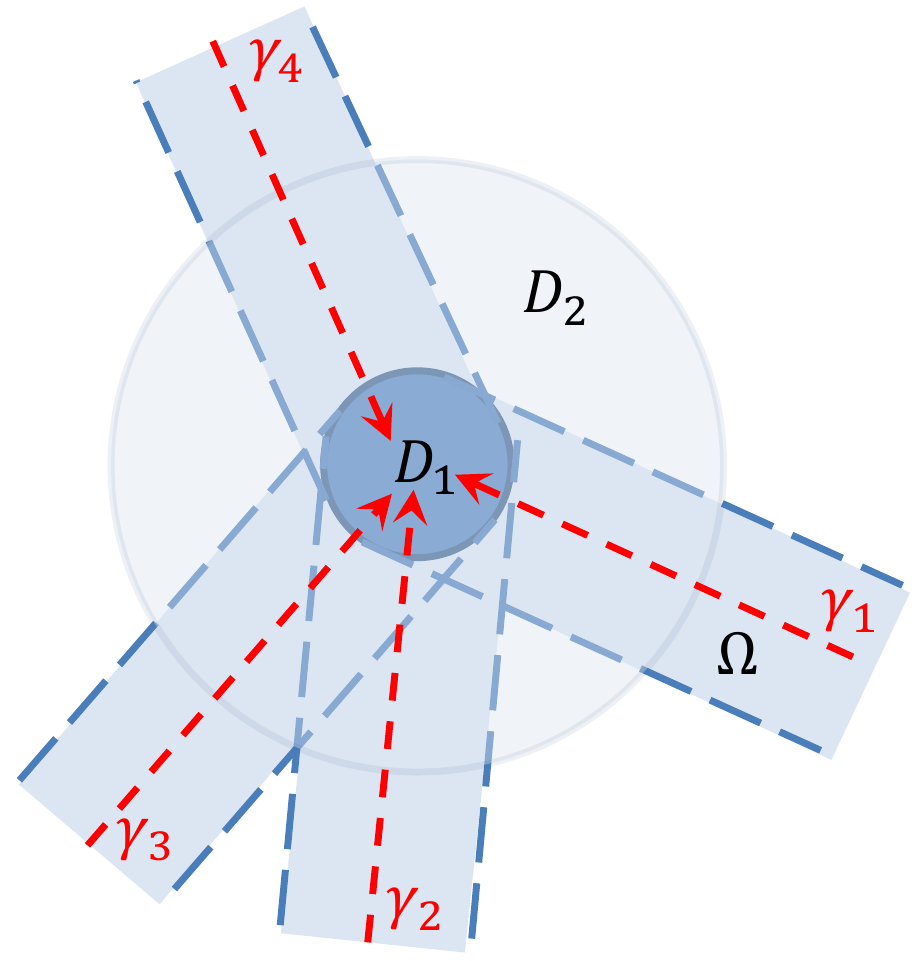}
\caption{If $\operatorname{supp}f\subseteq D_1$, then $\operatorname{supp}\mathcal{S}f\subseteq\Omega$\\ and $\mathcal{S}f|_{\partial D_2\cap\Omega}$ uniquely defines $\mathcal{S}f$ in\\ $\mathbb{R}^2\backslash D_2$. } \label{fig1b}
\end{subfigure}
\caption{(The color version is available online.) A star and the support of the corresponding transform.}
\label{fig1}
\end{figure}

In particular, if $c_1=\ldots=c_m=1$, the star transform $\mathcal{S}$ puts into correspondence to a given function $f(x)$ its integrals $\mathcal{S}f(x)$ along a two-parameter family of so-called ``stars''. Each star consists of a union of rays emanating from a common vertex $x$ along fixed directions $\gamma_i,\, i=1,\ldots,m$ (see Figure \ref{fig1a}).

\begin{definition}\label{def:aperture}
 Let the ray directions of the star be parameterized as
 \begin{equation}
    \gamma_j=(a_j,b_j)\in \mathbb{R}^2,\;\;j=1,\ldots,m.
 \end{equation}
 We call
 \begin{equation}
     a=(a_1,\ldots,a_m) \mbox{ and  }\; b=(b_1,\ldots,b_m)
 \end{equation}
 the \textbf{aperture vectors} of that star.
 Since $\gamma_j$'s are unit vectors
 \begin{equation}
     a_j^2+b_j^2=1,\;\;j=1,\ldots,m.
 \end{equation}
\end{definition}
Every star transform corresponds to an $m$-gon (polygon with $m$ sides) with vertices $c_1^{-1}\gamma_1,\dots , c_m^{-1}\gamma_m$. Many properties of the star transform can be characterised by geometric features of the corresponding polygon. The following special star plays a prominent role in some of our proofs.

\begin{definition}
 We call a star transform \textbf{symmetric}, if
 $m=2k$ for some $k\in\mathbb{N}$ and (after possible re-indexing) $\gamma_i=-\gamma_{k+i}$  with $c_i=c_{k+i}$ for all $i=1,\ldots, k$. The corresponding star is also called \textbf{symmetric}.
\end{definition}

\begin{definition}
 We call a star transform \textbf{regular}, if
 $c_1=\ldots=c_m=1$ and the ray directions $\gamma_i$, $i=1,\ldots,m$ correspond to the radius vectors of the vertices of a regular $m$-gon (see Figure \ref{fig4a}). The corresponding star is also called \textbf{regular}.
\end{definition}

We will show that the symmetric star transforms are the only non-invertible configurations of $\mathcal{S}$, while the regular star transforms with an odd number of vertices have the most stable inversions.

For regular stars with $m=2k+1$ rays, we parameterize the ray directions as follows (see Figure \ref{fig4a}):
\begin{equation}\label{sym-gamma}
    \gamma_j=(a_j,b_j)=(\cos\alpha_j, \sin\alpha_j),
\end{equation}
where
\begin{equation}\label{sym-alpha}
\alpha_1=0,\;\;\alpha_{2j}=\frac{2\pi j}{2k+1} \mbox{ and }
    \alpha_{2j+1}=-\frac{2\pi j}{2k+1},\;j=1,\ldots, k.
\end{equation}

\vspace{3mm}
Since $a_{2j}=a_{2j+1}$ and $b_{2j}=-b_{2j+1}$ for $j=1,\ldots,k$, the aperture vectors of the regular star with $2k+1$ rays can be expressed as
\begin{equation}\label{sym-ab}
    a=(1,a_2,a_2,\ldots,a_{2k}, a_{2k}),\;\;\;\;
    b=(0,b_2,-b_2,\ldots,b_{2k},-b_{2k}).
\end{equation}


\vspace{1mm}

Notice that if aperture vectors are fixed, the star is uniquely defined by its vertex $x$.
If one does not fix the directions of rays and allows arbitrary locations of its vertices, then the star transform will have $m+2$ degrees of freedom. The problem of inverting such a transform will be overdetermined.
We are interested in the inversion of the two-dimensional star transform $\mathcal{S}$ with $m$ rays in fixed directions $\gamma_i$ and non-zero weights $c_i,\, i=1,\ldots,m$.


\section{An Exact Inversion of the Star Transform}\label{sec:inversion}

Let us assume that  $\operatorname{supp}f\subseteq D_1$, where $D_1$ is an open disc of radius $r_1$ centered at the origin. Then $\mathcal{S}f$ is supported in an unbounded domain $\Omega$ (see Figure \ref{fig1b}). However, there exists a closed disc $\overline{D_2}$ of some finite radius $r_2>r_1$ centered at the origin such that $\mathcal{S}f$ is constant along the directions of rays $\gamma_i$ inside the corresponding ``strips'' of $\mathbb{R}^2\backslash \overline{D_2}$. In other words, the restriction of $\mathcal{S}f$ to $\overline{D_2}$ completely defines it everywhere else. Throughout the rest of the text we will assume that $\mathcal{S}f(x)$ is known for all $x\in \overline{D_2}$.


In this section we formulate and prove an exact closed-form inversion formula for $\mathcal{S}$ with arbitrary geometry and arbitrary non-zero weights. The intuitive idea of this inversion is the following. The Radon transform $\mathcal{R}$ of $g = \mathcal{S}f$ at $(\psi,{t})$ is the integral of the star transform data $g$ along the line $l(\psi,{t}) = \{x\in \mathbb{R}^2 |\, \langle x,\psi \rangle = {t} \}$ with normal vector $\psi$  and distance ${t}$ from the origin. For a generic line, this gives a sum of weighted integrals of $f$ over half-planes in $\mathbb{R}^2$, with different weights on different sides of $l(\psi,{t})$. By differentiating that quantity with respect to variable ${t}$, we recover the Radon data of $f$ along $l(\psi,{t})$. The inversion of $\mathcal{S}$ is then accomplished by inverting the Radon transform.\\

Before proving the theorem we discuss two auxiliary statements. They are geometric in nature and help with outlining the main ideas in the proof of Theorem \ref{Th:StarInversion}.

\subsection{Half Plane Transform and Its Derivative}

For a unit vector $\psi\in\mathbb{R}^2$ and ${t} \in \mathbb{R}$ define the half plane transform of $f$ as
\begin{equation}\label{def:half-plane}
 F_{\psi}({t}) = \int\limits_{\langle x, \psi \rangle \leq {t}} f\, dx.
\end{equation}

\begin{figure}[h]
\begin{center}
\includegraphics[height=1.5in]{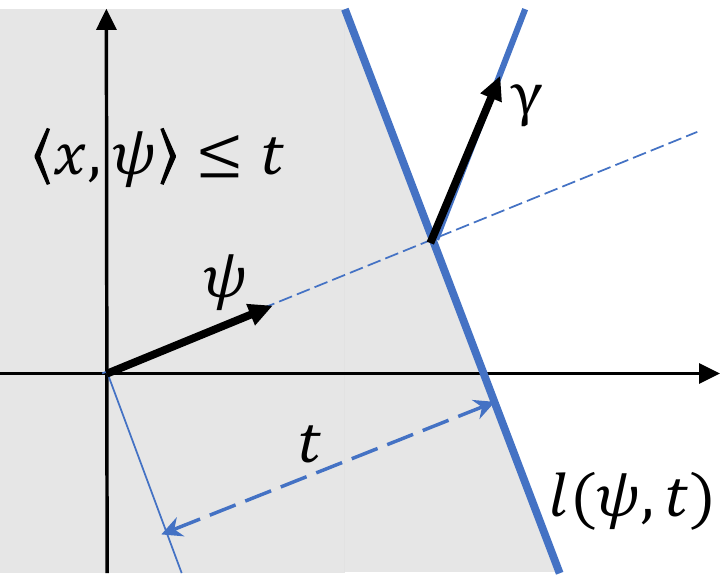}
\caption{A setup for the half-plane, divergent beam and Radon transforms.} \label{fig2}
\end{center}
\end{figure}

Geometrically,  $F_\psi({t})$ is the integral of $f$ over the half plane on one side of the line $l(\psi,{t})$ (see Fig. \ref{fig2}).

Now, let's look at the derivative of $F$ as a function of ${t}$
 \[
 \frac{dF_\psi({t}) }{d{t}} = \lim\limits_{h\rightarrow 0} \frac{ F_\psi({t} +h) - F_\psi({t}) }{h}.
 \]
Note that $ F_\psi({t} +h) - F_\psi({t})$ is the integral of $f$ over the infinite strip between the lines  $l(\psi,{t})$ and  $l(\psi,{t}+h)$, which plays a central role in our construction.
Using Fubini's theorem and the fundamental theorem of calculus, we get

\begin{align*}
\frac{dF_\psi({t}) }{d{t}}  &= \frac{d}{d{t}} \int\limits_{\langle x, \psi \rangle \leq {t}} f\, dx
 = \frac{d}{d{t}} \int_{-\infty}^{t} \mathcal{R}f(\psi,{s})\, d{s}
 = \mathcal{R}f(\psi,{t}),
\end{align*}
where $\mathcal{R}$ is the standard Radon transform.  Thus we have proved the following

 \begin{lemma} \label{ftcradon}
 	For function $f \in C_c{({D_1})}$ we have
 	\begin{equation}
 	  \frac{dF_\psi({t}) }{d{t}} = \mathcal{R}f(\psi,{t}).
 	\end{equation}
 \end{lemma}

A key step in our proof is the possibility of expressing the half plane transform $F_\psi$ of $f$ in terms of its star transform $\mathcal{S}f$. We show this by first finding a relation between $F_\psi$ and $\mathcal{X}_\gamma(f)$.

If $\langle \psi,\gamma \rangle \neq 0$ we can integrate $\mathcal{X}_\gamma(x)$ along the line  $l(\psi,{t})$ to get an expression in terms of $F_{\psi}({t})$. Namely

\begin{lemma} \label{radonray}
Let $f \in C_c({D_1})$ and $\langle \psi,\gamma \rangle \neq 0 $. Then
\begin{equation}
\frac{d}{d{t}}\mathcal{R}(\mathcal{X}_\gamma f)(\psi,{t}) = \frac{-1}{\langle \psi,\gamma\rangle} \frac{dF_{\psi}({t})}{d{t}}.
\end{equation}
\end{lemma}
\noindent \textbf{Proof.} Depending on whether $\langle \psi,\gamma \rangle$ is positive or negative, $\mathcal{R}(\mathcal{X}_\gamma f)(\psi,{t})$ is integrating $f$ with the weight $\langle \psi,\gamma \rangle^{-1}$ over the half plane on one or the other side of the line $l(\psi,{t})$ (see Figure \ref{fig2}). This is an implication of a modified version of Fubini's theorem (e.g. see the Moving Sections Lemma in \cite{Amb-Latifi1}).   If $\langle \psi,\gamma \rangle > 0 $ we have
\[
\mathcal{R}(\mathcal{X}_\gamma f)(\psi,{t}) =  \frac{1}{\langle \psi,\gamma\rangle}  \left[  \int_{\mathbb{R}^2}f\, dx -  F_{\psi}({t}) \right]
\]
and if $\langle \psi,\gamma \rangle < 0 $
\[
\mathcal{R}(\mathcal{X}_\gamma f)(\psi,{t}) = \frac{-1}{\langle \psi,\gamma\rangle} F_{\psi}({t}).
\]
In either case, taking the derivative of both sides yields the same expression and finishes the proof. $\hfill\blacksquare$ \\


\subsection{Main Result and Some Corollaries}

{
\begin{definition}\label{def:sing-dir}
 Let $\mathcal{S}$ be the star transform defined in (\ref{def:Star}). We call
 \begin{equation}\label{eq:Z1}
     \mathcal{Z}_1=\bigcup\limits_{j=1}^m\{\psi:\langle\psi,\gamma_j\rangle=0\}
 \end{equation}
 the set of singular directions of Type 1 for $\mathcal{S}$, and
 \begin{equation}\label{eq:Z2}
      \mathcal{Z}_2=\left\{\psi:
     \sum_{j=1}^m c_j \prod_{i \neq j}\langle\psi,\gamma_j\rangle=0\right\}
 \end{equation}
 the set of singular directions of Type 2 for $\mathcal{S}$. Now define
	\begin{equation}\label{def:q}
	w(\psi)=\displaystyle\sum_{i=1}^{m}\frac{c_i}{\langle \psi, \gamma_i \rangle},\;\;\psi\in \mathbb{S}^1\setminus\mathcal{Z}_1,\;\;\;\;\;
	q(\psi) = \frac{-1}{w(\psi)},\;\;\psi\in \mathbb{S}^1\setminus (\mathcal{Z}_1\cup\mathcal{Z}_2).
	\end{equation}
\end{definition}

\begin{theorem} \label{Th:StarInversion}
    Let $f \in C_c(D_1)$, $\mathcal{S}$
	be the star transform and $\mathcal{R}$ be the Radon transform of a function in $\mathbb{R}^2$.
	 Then for any $\psi\in \mathbb{S}^1\setminus (\mathcal{Z}_1\cup\mathcal{Z}_2)$ we have:
	\begin{equation}\label{eq:Rad-Star}
	\mathcal{R}f(\psi,{t}) = q(\psi) \frac{d}{d{t}}\mathcal{R}(\mathcal{S}f)(\psi,{t}).
	\end{equation}
\end{theorem}
}

 \noindent \textbf{Proof.} Using Lemmas  \ref{ftcradon} and \ref{radonray} we get
\[
\frac{d}{d{t}}\mathcal{R}(\mathcal{X}_\gamma f)(\psi,{t})  = \frac{-1}{\langle \psi,\gamma\rangle} \mathcal{R}f(\psi,{t}).
\]
Now we write $\mathcal{S}$ in terms of $\mathcal{X}_{\gamma_i}$'s
\begin{align*}
\frac{d}{d{t}}\mathcal{R}(\mathcal{S}f)(\psi,{t}) = &\frac{d}{d{t}} \mathcal{R}\left(\sum_{i=1}^{m}{c_i\,\mathcal{X}_{\gamma_i}f}\right)(\psi,{t})
= \sum_{i=1}^{m} {c_i\, \frac{d}{d{t}}\mathcal{R}(\mathcal{X}_{\gamma_i} f)(\psi,{t})} \\
= &\sum_{i=1}^{m} \frac{-c_i}{\langle \psi,\gamma_i\rangle}\; \mathcal{R}f(\psi,{t})\;  =-w(\psi)\;\mathcal{R}f(\psi,{t}),
\end{align*}
which ends the proof. $\hfill\blacksquare$ \\

It is easy to notice that if $\mathcal{Z}_1\cup\mathcal{Z}_2$ is finite, then the singularities appearing in the right hand side of formula (\ref{eq:Rad-Star}) are removable. Namely,
\begin{remark}
 Let $\mathcal{Z}_1\cup\mathcal{Z}_2=\left\{\zeta_i\right\}_{i=1}^M$. Then by formula \eqref{eq:Rad-Star} and continuity of $\mathcal{R}f$ we have
\begin{equation}
 \lim_{\psi\to\zeta_i}\left[q(\psi) \frac{d}{d{t}}\mathcal{R}(\mathcal{S}f)(\psi,{t})\right]=\mathcal{R}f(\zeta_i,{t})<\infty.
\end{equation}
\end{remark}
In other words, the full Radon data can be recovered from the star transform.

\begin{corollary}
 	If $\mathcal{Z}_1\cup\mathcal{Z}_2$ is finite, we can apply $\mathcal{R}^{-1}$ to recover $f$.
\end{corollary}

 We finish this section with statements of two special cases of Theorem \ref{Th:StarInversion} and discussion of some pertinent relations. In the case of $m=2$, Theorem \ref{Th:StarInversion} yields a new and simple inversion for the V-line transform.

\begin{corollary}\label{cor-vline}
An inversion formula for the V-line transform $\mathcal{V} = c_1\mathcal{X}_{\gamma_1}+c_2\mathcal{X}_{\gamma_2}$ with fixed non-collinear ray directions $\gamma_1,\,\gamma_2$ and nonzero weights $c_1,\, c_2$ is given by
\begin{equation}
f = \mathcal{R}^{-1}\left[ \frac{-\langle \psi, \gamma_1 \rangle \langle \psi, \gamma_2 \rangle }{c_2\langle \psi, \gamma_1 \rangle + c_1\langle \psi, \gamma_2 \rangle } \frac{d}{d{t}}\mathcal{R}(\mathcal{V}f)(\psi,{t})\right].
\end{equation}
\end{corollary}

\vspace{3mm}

A curious special case of Theorem \ref{Th:StarInversion} is the following.

\begin{corollary}\label{cor-div_beam}
An inversion of the divergent beam transform $\mathcal{X_\gamma}$ is given by the formula
\begin{equation}
f = \mathcal{R}^{-1}\left[ -\langle \psi,\gamma \rangle \frac{d}{ d{t}} \mathcal{R}(\mathcal{X_\gamma}f)(\psi,{t})\right].
\end{equation}
\end{corollary}

The directional derivative  $D_{-\gamma}$, given by $ D_{\gamma}h\,(x)=\gamma\cdot\nabla h\,(x)$, is the natural inverse of $\mathcal{X_{\gamma}}$. Hence, for any $h\in U_\gamma=\left\{\mathcal{X}_\gamma(f):\,f\in C_c(\mathbb{R}^2)\right\}$ in the range of $\mathcal{X}_\gamma$, one can write the following identity
\begin{equation}\label{eq:divbeaminv}
     D_\gamma h\,(x) = \mathcal{R}^{-1}\left[\langle \psi,\gamma \rangle \frac{d}{d{t}} \mathcal{R}\right] h\,(x).
\end{equation}

The identity (\ref{eq:divbeaminv}) is not really new. It is equivalent to a known relation
\begin{equation}\label{eq:RPD}
P(D_{e_1} , D_{e_2}) = \mathcal{R}^{-1}\,\left[P\left(\psi_1 \frac{d}{dt}, \psi_2 \frac{d}{dt}\right) \right]\,\mathcal{R},
\end{equation}
where $e_i$'s are standard orthonormal vectors and $P$ is a given polynomial of two variables. One can get (\ref{eq:RPD}) by repeatedly applying (\ref{eq:divbeaminv}) and canceling $\mathcal{R}^{-1}\mathcal{R}$ in the resulting telescoping identity. The other direction of equivalence is trivial.


\section{Injectivity of the Star Transform}\label{sec:inj}

In certain configurations of the star transform, the function $q(\psi)$ is not defined for any $\psi$. For example, this happens when $m=2$, $c_1=c_2=1$, and $\gamma_1=-\gamma_2=(1,0)$. In this case $\langle \psi,\gamma_1\rangle+\langle\psi,\gamma_2\rangle\equiv0$ for any $\psi\in \mathbb{S}^1$.

Consider a similar configuration with more rays: $m=4$, $c_1=\ldots=c_4=1$, $\gamma_1=-\gamma_3$ and $\gamma_2=-\gamma_4$. While it is not obvious that $q(\psi)$ is not defined at any $\psi$, it is easy to notice that in this case the star transform is not injective, since it provides less information than the ordinary Radon transform restricted  to two directions. Interestingly enough, such configurations are the only ones, for which the star transform is not injective.

\begin{theorem}\label{th:injectivity}
The star transform $\mathcal{S} = \sum_{i=1}^m c_i\mathcal{X}_{\gamma_i}$
is  invertible \textbf{if and only if}
it is not symmetric.
\end{theorem}

An important object in our inversion is the function $q(\psi)$ (or equivalently its reciprocal $w(\psi)$). It leads to the geometric condition that is necessary for the star transform to be non-invertible. To describe the connection between $q(\psi)$ and the geometric condition we need to consider the elementary symmetric polynomial of degree $m-1$ in $m$ variables $y=(y_1,\ldots,y_m)$ (see \cite{McDonald})
\begin{equation}\label{def:elem_poly}
e_{m-1}(y_1,\dots,y_m)= \sum_{i=1}^m \prod_{j \neq i} y_j.
\end{equation}
 Using the above notation, one can re-write formula (\ref{def:q}) as
\begin{equation}\label{def_q2}
q(\psi) = \frac{c_1^{-1}\ldots c_m^{-1}}{\displaystyle\frac{e_{m-1}\left(\langle \psi, c_1^{-1}\gamma_1 \rangle, \dots ,\langle \psi, c_m^{-1}\gamma_m \rangle\right)}{\langle \psi, \gamma_1 \rangle \dots \langle \psi, \gamma_m \rangle}}.
\end{equation}

\vspace{2mm}

\noindent The proof of Theorem \ref{th:injectivity} is based on the description of zero sets of $e_{m-1}$ and the fact that the star transform $\mathcal{S}$ is invertible if $e_{m-1}(\langle \psi, c_1^{-1}\gamma_1 \rangle, \dots ,\langle \psi, c_m^{-1}\gamma_m \rangle)$ is not identically zero as a function of $\psi$.\\

If $\hat{\psi}=(r,s)\in \mathbb{R}^2 $ (not necessarily on $\mathbb{S}^1$) and $c\doteq c_1\ldots c_m$, we define a polynomial in two variables $(r,s)$ as follows:
\begin{equation}\label{def:poly-psi}
  P_2(\hat{\psi})=c\, e_{m-1}(\langle \hat{\psi}, c_1^{-1}\gamma_1 \rangle, \dots ,\langle \hat{\psi}, c_m^{-1}\gamma_m \rangle).
\end{equation}

\begin{lemma}\label{zeros_of_P2}
The polynomial $P_2(r,s)$ either has finitely many zeros on $\mathbb{S}^1$ or it is identically zero.
\end{lemma}
\noindent \textbf{Proof.} Let us start with noticing that $e_{m-1}(y)$ is a homogeneous polynomial of degree $d-1$, i.e. $e_{m-1}(\lambda y)=\lambda^{m-1}e_{m-1}(y)$.
Hence, $P_2(r,s)$ is also a homogeneous polynomial of degree $m-1$, which either has finitely many (projective) roots or is identically zero.
$\hfill\blacksquare$ \\

\noindent \textbf{Proof (of Theorem \ref{th:injectivity}).} Assume that for a fixed choice of  $\gamma_1, \dots, \gamma_m$ there is no inversion for the corresponding star transform. By Theorem \ref{Th:StarInversion}, $P_2(\psi)$ has to be zero on an infinite subset of $\mathbb{S}^1$. Therefore, by Lemma \ref{zeros_of_P2}, $P_2(\psi)\equiv0$.

Using the notation introduced in (\ref{def:q})
$$
w(\psi)=\displaystyle\sum_{i=1}^{m}\frac{c_i}{\langle \psi, \gamma_i \rangle}=\frac{P_2(\psi)}{\prod\limits_{i=1}^m \langle\psi,\gamma_i\rangle}.
$$
If $w(\psi)=0$ away from the poles, then
    $$
    0=\lim\limits_{\psi\to\gamma_j^{\perp}}\langle\psi,\gamma_j\rangle w(\psi).
    $$
    This is $c_j$ if there is no other $\gamma_i$ parallel to $\gamma_j$ and is equal to $c_j-c_{j'}$ if $\gamma_{j'}=-\gamma_{j}$. The first case can not hold since $c_j\ne0$ so the second must hold, which implies that $\mathcal{S}$ is symmetric.  $\hfill\blacksquare$

 Theorem \ref{th:injectivity} immediately implies the following.
\begin{corollary}
 Any star transform with an odd number of rays is invertible.
\end{corollary}


\section{Singular Directions of the Star Transform}\label{sec:sing-dir}

 The number and location of singular directions affect the quality of numerical reconstructions. The singular directions of Type 2 correspond to ``division by zero'' of the processed data  $\frac{d}{d{t}}\mathcal{R}(\mathcal{S}f)$, while those of Type 1 correspond to ``multiplication by zero''. Hence, it is natural to expect that singular directions of Type 2 will create instability and adversely impact the reconstruction. Our numerical experiments confirm these expectations (see Section \ref{sec:numerical}). It is also interesting, that the (totally different) algorithm for inversion of the star transform obtained in \cite{ZSM-star-14} produces a relation equivalent to the defining relation of $\mathcal{Z}_2$ in (\ref{eq:Z2}) as a necessary and sufficient condition for the instability of that algorithm (see formula (51) on page 18 in \cite{ZSM-star-14}).

While the geometric meaning of singular directions of Type 1 is obvious for any $m$, there is no easy interpretation of set $\mathcal{Z}_2$ for $m\ge 3$.  However, the singular directions of Type 2 are more crucial for the quality of reconstruction.

In Section \ref{sec:T2-unif} we discuss the existence of singular directions of Type 2 in star transforms with uniform weights. In Section \ref{Sec:reg-star} we show the absence of Type 2 singularities in regular star transforms with an odd number of rays. The case of star transforms with non-uniform weights is discussed in Section \ref{sec:T2-nonunif}.

\subsection{Star Transforms with Uniform Weights}\label{sec:T2-unif}

\begin{theorem}\label{th:singularities} Consider the star transform $\mathcal{S} = \sum_{i=1}^m \mathcal{X}_{\gamma_i}$ with uniform weights.
\begin{enumerate}
    \item If $m$ is even, $\mathcal{S}$ must contain a singular direction of Type 2.
    \item When $m$ is odd, there exist configurations of $\mathcal{S}$ that contain singular directions of Type 2, as well as configurations that do not contain them.
\end{enumerate}
\end{theorem}
\begin{figure}[ht]
\begin{subfigure}{.5\textwidth}
  \centering
 \includegraphics[height=1.4in]{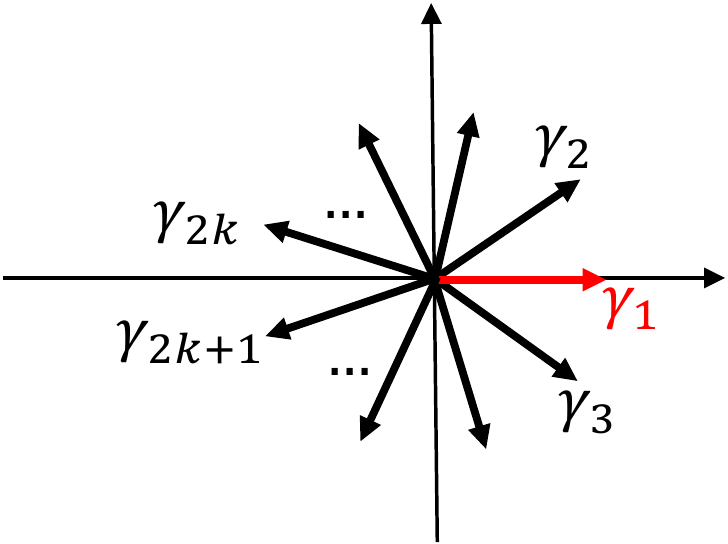}
 \caption{A star with $2k+1$ symmetric rays and without a singular direction of Type 2.}
  \label{fig4a}
\end{subfigure}
\begin{subfigure}{.5\textwidth}
  \centering
\includegraphics[height=1.4in]{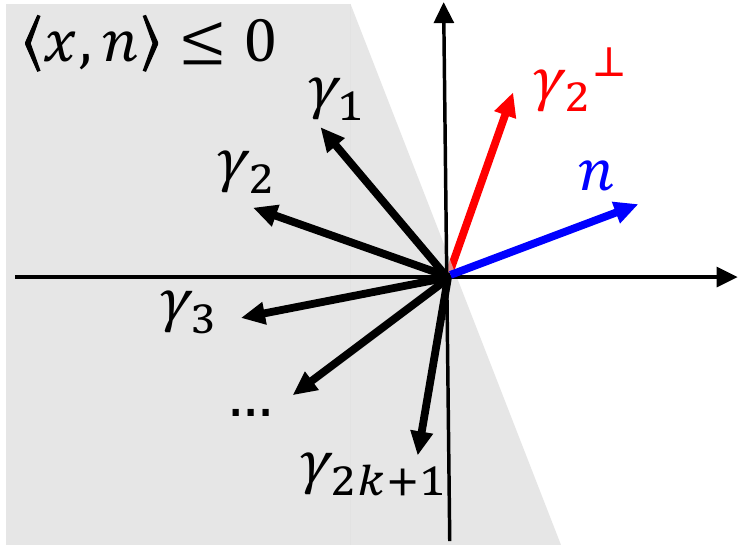}
\caption{A star with $2k+1$ rays that has a singular direction of Type 2.} \label{fig4b}
\end{subfigure}
\caption{Stars of $2k+1$ rays with and without singular directions of Type 2.}
\label{fig4}
\end{figure}

\vspace{2mm}

\noindent \textbf{Proof (of Part 1).}
Let $m=2k$. Then polynomial $P_2$ defined in (\ref{def:poly-psi}) is an odd function on the unit circle, i.e.
$$
  P_2(-\psi)=-P_2(\psi), \;\; \psi\in\mathbb{S}^1.
$$
If $P_2\not\equiv 0$, then there exists $\psi^*\in\mathbb{S}^1$ such that $P_2(-\psi^*)=-P_2(\psi^*)>0$. Hence, by continuity of $P_2$ it has to be zero on each arc of $\mathbb{S}^1$ between $\psi^*$ and $-\psi^*$. $\hfill\blacksquare$


\vspace{5mm}


\noindent \textbf{Proof (of Part 2a).}

\noindent Now let us show that for an \textbf{odd number of rays}, there exist star transforms \textbf{with singularities of Type 2}. In other words, for any $m=2k+1$, $k\in\mathbb{N}$ there exist configurations of ray directions $\gamma_i$, $i=1,\ldots,m$, such that $P_2(\psi)=0$ for some $\psi\in \mathbb{S}^1$, where $P_2$ is the polynomial  defined in (\ref{def:poly-psi}).

A trivial example is when a pair of rays point to opposite directions, i.e. $\gamma_i=-\gamma_j$ for some indices $i$ and $j$. Then it is easy to see that the direction normal to them is a singular direction of Type 2. A more interesting example is discussed below.

Consider an arbitrary open half-plane $H=\left\{x: \langle x, n\rangle< 0\right\}$, where $n$ is the outward normal to its boundary passing through the origin.
Choose a set of distinct ray directions $\left\{\gamma_i\right\}_{i=1}^m$ so that $\gamma_i\in H$ 
for all $i=1,...,m$. Without loss of generality, let us assume that $\gamma_i$'s are indexed according to the growth of their polar angle (see Figure \ref{fig4b}).

 Consider $\psi=\gamma_2^{\perp}$ oriented so that $\langle \psi, \gamma_1\rangle>0$. Then $\langle \psi, \gamma_2\rangle=0$ and $\langle \psi, \gamma_i\rangle<0$ for all $i>2$.
Estimating the terms of the sum in (\ref{def:elem_poly}), it is easy to notice that  $\prod_{j \neq 2}\langle \psi, \gamma_j\rangle <0$, while all of the other terms are 0. Thus, for $\psi=\gamma_2^{\perp}$ we get $P_2(\psi)<0$.

On the other hand, if $\psi=n$, then all terms of the sum in  (\ref{def:elem_poly}) are positive and $P_2(\psi)>0$. By continuity of $P_2(\psi)$, there has to exist a value of $\psi$, for which $P_2(\psi)=0$. $\hfill\blacksquare$


\vspace{5mm}

\noindent \textbf{Proof (of Part 2b).}
The \textbf{regular} star transform $\mathcal{S}$ with $m=2k+1$ rays does not have a singular direction of Type 2. See Section \ref{Sec:reg-star}. $\hfill\blacksquare$

\subsection{Regular Star Transforms with an Odd Number of Rays}\label{Sec:reg-star}
In this section we show that any regular star transform with an odd number of rays (see Figure \ref{fig4a}) has a stable inversion. More specifically,
\begin{theorem}{\label{th:reg}}
The regular star transform $\mathcal{S}$ with $m=2k+1$ rays does not have a singular direction of Type 2.
\end{theorem}

\noindent \textbf{Proof.} We need to show that for the regular star with $m=2k+1$ rays, $P_2(\psi)\ne0$ for any $\psi\in \mathbb{S}^1$, where $P_2$ is the polynomial  defined in (\ref{def:poly-psi}).

Using the parameterization  $\psi(\alpha)=(\cos\alpha,\sin\alpha)$, we denote
\begin{equation}
   F(\alpha)=P_2(\psi(\alpha)).
\end{equation}
We will show that $F(\alpha)\ne0$ for any $\alpha\in[0,2\pi]$. Our argument is based on the following statement, proved after the proof of Theorem \ref{th:reg}.

\begin{lemma}\label{lem-P1}
There exists  a polynomial of one variable $P_1$ of degree $\le m-1$, such that
\begin{equation}
    F(\alpha)=P_1(\cos\alpha),
\end{equation}
\end{lemma}

   Due to the symmetry of $P_2(\psi)$ with respect to $\gamma_i$'s and the symmetric distributions of $\gamma_i$'s on $\mathbb{S}^1$, it follows that $F(\alpha)$ is periodic with period $2\pi/m$. Therefore all of its Fourier coefficients $F_k$ are zero unless $k$ is divisible by $m$. Since $F$ is a trigonometric polynomial, it is equal to its Fourier expansion, and since it is of degree at most $(m-1)$, the only index in its Fourier expansion divisible by $m$ is $k = 0$, thus $F$ is constant. By Theorem
   \ref{th:injectivity} that constant can not be zero, which completes the proof of Theorem \ref{th:reg}. $\hfill\blacksquare$
\\

\noindent \textbf{Proof of Lemma \ref{lem-P1}:}
Recall Definition \ref{def:aperture} and Equation (\ref{sym-ab}). Using our previous notations $\psi(\alpha)=(\cos\alpha,\sin\alpha)\doteq (r,s)$ we can write
\begin{align*}
    F(\alpha)&=P_2(\psi(\alpha))=e_{m-1}(\langle \psi, \gamma_1 \rangle, \dots ,\langle \psi, \gamma_m \rangle)=e_{m-1}(ra+sb)\\
    &= \sum_{i=1}^m \prod_{j \neq i} (r a_j + s b_j)
    = \prod_{j \neq 1} (r a_j + s b_j)+\left[\prod_{j \neq 2} (r a_j + s b_j)+\prod_{j \neq 3} (r a_j + s b_j)\right]\\
    &+\ldots+ \left[\prod_{j \neq 2k} (r a_j + s b_j)+\prod_{j \neq 2k+1} (r a_j + s b_j)\right]
\end{align*}

It easy to notice, that after appropriate cancellations inside the brackets using (\ref{sym-ab}), the last expression does not contain any odd degree of $s$.
Substituting $s^2=1-r^2$ we obtain the desired result. $\hfill\blacksquare$\\

In the case of $m=3$, there is a simpler proof of Theorem \ref{th:reg}. That proof also shows that the transform $\mathcal{S}$ corresponding to the regular star has (in some sense) the most stable inversion. It is based on the following result from \cite{Conflitti}:

\noindent\textbf{Theorem} \textsl{The zero set of $e_2(y_1,y_2,y_3)$ is the circular cone
\begin{equation}\label{cone}
    e_2^{-1}(0)=\left\{y\in\mathbb{R}^3:\,\cos^2{(y,u)}=\frac{1}{3}\right\},
\end{equation}
where $u=(1,1,1)$ and $(y,u)$ denotes the angle between vectors $y$ and $u$.}

\vspace{3mm}

\noindent\textbf{An Alternative Proof of Theorem \ref{th:reg} when  $m=3$.}

\noindent We need to show that $e_2(ra+sb)\ne0$, where $a$ and $b$ are the aperture vectors of the star transform defined by (\ref{sym-gamma}), (\ref{sym-alpha}), (\ref{sym-ab}) as
\begin{equation}\label{r3-ap}
a=\left(1, -\frac{1}{2}, -\frac{1}{2}\right), \;\; b=\left(0, \frac{\sqrt3}{2}, -\frac{\sqrt3}{2}\right),
\end{equation}
and $r^2+s^2=1$. In light of (\ref{cone}), it is equivalent to proving that the plane $T$ spanned by aperture vectors $a$ and $b$ intersects the circular cone defined in (\ref{cone}) only at the origin. Since
$$
\langle a,u \rangle =0 \mbox{ and } \langle b,u \rangle =0,
$$
it is clear that $u\perp T$. Hence,  $T\bigcap e^{-1}_2(0)=\{0\}$. Moreover, the aperture vectors $a$ and $b$ corresponding to the regular star, span the plane that is ``as far as possible'' from the zero set of $e_2$.
$\hfill\blacksquare$


\subsection{Star Transforms with Non-Uniform Weights}\label{sec:T2-nonunif}

 The regular stars with an odd number of rays are not the only configurations of the transform $\mathcal{S}$  that do not have a singular direction of Type 2. Due to continuous dependence of $P_2(\psi)$ on $\alpha_i$'s, small perturbations of $\gamma_i$'s from symmetry positions will not introduce a  singular direction of Type 2. Similarly, small perturbations of uniform weights $c_1=\ldots=c_m=1$ to non-uniform values will not introduce singular directions of Type 2. Moreover,

\begin{remark}\label{rem:sign-pos}
 The polynomial $P_2(\psi)$ does not change if one simultaneously replaces $\gamma_i$ by $-\gamma_i$ and $c_i$ by $-c_i$ for any $i$. This provides a recipe for creating star transforms with weights of mixed algebraic signs that have stable inversions.
\end{remark}

For example, modifying the regular star transform of 3 rays (defined by formula (\ref{r3-ap})) through replacement of $\gamma_2$ by $-\gamma_2$ and changing the sign of $c_2=1$, we get a stable configuration with weights $c_1=1$, $c_2=-1$, $c_3=1$ and aperture vectors:
$$
a=\left(1, \frac{1}{2}, -\frac{1}{2}\right), \;\; b=\left(0, -\frac{\sqrt3}{2}, -\frac{\sqrt3}{2}\right).
$$
In general, one can get stable setups when the absolute values of weights are close to each other.

The star transforms with weights of mixed algebraic signs satisfying the condition
\begin{equation}
    \sum_{i=1}^mc_i=0
\end{equation}
play an important role in simultaneous reconstruction of scattering and absorption coefficients in single scattering tomography \cite{Florescu2, Florescu3, Kats-Kryl-13, ZSM-star-14}. Simultaneously satisfying the above condition and keeping the absolute values of weights as close to each other as possible, suggest choosing the weights (up to a proportionality coefficient and re-indexing) as follows:
\begin{equation}
    c_1=\ldots=c_k=-\frac{1}{k},\;\;c_{k+1}=\ldots=c_{2k+1}=\frac{1}{k+1},\;\;m=2k+1.
\end{equation}
In the case of $m=3$, this implies choosing the weights proportional to $c_1=c_2=1$ and $c_3=-2$. Such a setup was analyzed in detail in \cite{ZSM-star-14}.

\vspace{5mm}

We finish this section with generalizations of the statements of Section \ref{sec:T2-unif} to the case of transforms with non-uniform weights. For the stars with an even number of rays, both the result and its proof are almost identical. Namely,
\begin{theorem}\label{th:sing-gen-even} Consider the star transform $\mathcal{S} = \sum_{i=1}^m c_i\mathcal{X}_{\gamma_i}$. If $m$ is even, $\mathcal{S}$ must contain a singular direction of Type 2.
\end{theorem}
In the case of star transforms with an odd number of rays, it should be clarified in what sense we want to generalize Part 2 of Theorem \ref{th:singularities}. If one is allowed to choose both the weights $c_i$ and the ray directions $\gamma_i$ to claim existence of setups with and without Type 2 singularities, then the statement has a trivial proof by choosing uniform weights and referring to Theorem \ref{th:singularities}. Hence, we consider the following question. Are there configurations (i.e. ray directions $\gamma_1,\ldots,\gamma_m$) with/without Type 2 singularities for a \textbf{specified}
set of weights $c_1, \ldots, c_m$?

\begin{theorem}\label{th:sing-gen-odd} Consider a star transform $\mathcal{S} = \sum_{i=1}^m c_i\mathcal{X}_{\gamma_i}$, where $m=2k+1$.

\vspace{2mm}

\noindent (a) For any set of specified weights $c_1 \ldots, c_m$, there exist $\gamma_1,\ldots,\gamma_m$ such that $\mathcal{S}$ contains singular directions of Type 2. In other words, for any set of weights there are configurations of $\mathcal{S}$
with unstable inversion.
\vspace{2mm}

\noindent (b) Let $m=3$. For any set of specified weights $c_1, c_2, c_3$, there exist $\gamma_1,\gamma_2,\gamma_3$ such that $\mathcal{S}$ does not contain singular directions of Type 2. In other words, for any set of weights there are configurations of $\mathcal{S}$ with a stable inversion.
\end{theorem}
\noindent \textbf{Proof of (a): Existence of configurations with Type 2 singularities.}\\ Both examples provided in the proof of Theorem \ref{th:singularities} work here with small modifications. If $\gamma_i=-\gamma_j$ for some indices $i$ and $j$, then the direction normal to them is a singular direction of Type 2.

A more interesting example is the case when all $\gamma_i$'s belong to the same half-plane. Recall the corresponding setup  from the proof of Theorem \ref{th:singularities}. Without loss of generality, let us assume that the $c_1c_2>0$, i.e. the weights $c_1$ and $c_2$ have the same sign. Since we consider the case of $m=2k+1$ for $k\ge1$, the assumption above is just a matter of re-indexing the weights (if necessary).

Let $\psi_1=\gamma_1^{\perp}$ and $\psi_2=\gamma_2^{\perp}$ be vectors obtained by a clockwise rotation by $\pi/2$ from $\gamma_1$ and $\gamma_2$ correspondingly. Then, it is easy to verify that $P_2(\psi_1)$ and $P_2(\psi_2)$ have different signs. Hence, by continuity of $P_2$ there exists some $\psi$ s.t. $P_2(\psi)=0$. $\hfill\blacksquare$ \\

Before proceeding to the proof of the second statement of Theorem \ref{th:sing-gen-odd}, we define a new class of planes in $\mathbb{R}^3$ and discuss their properties.

\begin{definition}
 We call a plane $T\subseteq\mathbb{R}^3$ ``spannable by star aperture vectors'' or SSAV if $T=\operatorname{span}\{a,b\}$, where $a,b\in\mathbb{R}^3$ are aperture vectors of some star with $m=3$ rays. In particular this means that $(a_i,b_i) $ are unit vectors in $\mathbb{R}^2$ for $i=1,2,3$.
\end{definition}

Not every plane (containing the origin) in $\mathbb{R}^3$ is SSAV. The following lemma gives a complete description of admissible normal vectors to an SSAV plane.

\begin{lemma}
A vector $n=(n_1,n_2,n_3)\in\mathbb{R}^3$ is normal to an SSAV plane if and only if $\left|n_1\right|$, $\left|n_2\right|$ and $\left|n_3\right|$ correspond to side lengths of some triangle. We will denote the set of all such vectors by $A$.
\end{lemma}

\noindent \textbf{Proof.} Let $n$ be normal to a plane $T\in\mathbb{R}^3$ containing the origin. $T$ is SSAV if and only if there exists a star with aperture vectors $a,b$ such that $\langle n,a\rangle=\langle n,b\rangle=0$. By definition of the aperture vectors, the last relation implies
$$
n_1\gamma_1+n_2\gamma_2+n_3\gamma_3=0,
$$
where $\gamma_1$, $\gamma_2$ and $\gamma_3$ are the ray directions of the corresponding star. Since $\gamma_i$'s are unit vectors, the statement of the lemma follows from the above equality interpreting it as a sum of vectors tracing the perimeter of a triangle.
$\hfill\blacksquare$

The set $A$ of admissible normals to an SSAV plane has an interesting geometric description due to the three triangle inequalities
$$
\left|n_1\right|+\left|n_2\right|\ge\left|n_3\right|,\quad
\left|n_2\right|+\left|n_3\right|\ge\left|n_1\right|,\quad
\left|n_1\right|+\left|n_3\right|\ge\left|n_2\right|.
$$
In the octant $n_1>0$, $n_2>0$, $n_3>0$ it coincides with an infinite polyhedral cone with a vertex at the origin and infinite edges along vectors $(0,1,1)$, $(1,0,1)$ and $(1,1,0)$. We denote the portion of $A$ in this first octant by $A^+$. In all other octants $A$ coincides with a similar tetrahedron with the edges along appropriately signed versions of the same vectors.

Formula (\ref{cone}) shows that every plane with a normal vector $n\in A^+$ intersects $e_2^{-1}(0)$ only at the origin. Moreover, among the SSAV planes, those with a normal vector in $A^+$ are the only ones that intersect $e_2^{-1}(0)$ only at the origin. Hence, the set $A^+$ contains normals to SSAV planes that correspond (through the spanning aperture vectors) to star transforms with uniform weights $c_1=c_2=c_3=1$ and a stable inversion, i.e. without a singular direction of Type 2.

Let $$W=\operatorname{diag}(c_1,c_2, c_3)$$ denote the diagonal matrix of weights.
\begin{lemma}\label{lem:plane-trans}
For any set of specified weights $c_1,c_2,c_3>0$ there exist  SSAV planes $T_1,T_2$ such that $T_2=W^{-1}T_1$ and $T_i\cap e_2^{-1}(0)=\{0\}$ for $i=1,2$.
\end{lemma}
\textbf{Proof.} Let $n^{(1)}, n^{(2)}$ denote some vectors normal to the planes $T_1$ and $T_2$ correspondingly. The proof will follow immediately if we show that for any $c_1,c_2, c_3>0$ there exist $n^{(1)}, n^{(2)}\in A^+$  such that $n^{(2)}=W n^{(1)}$.

We first note that the problem is invariant with respect to multiplication of all weights $c_1, c_2, c_3>0$ by a positive constant, as well as their permutation. Hence, without loss of generality  we will assume that $0<c_1\le c_2=1\le c_3$.

A choice of desired pair of vectors $n^{(1)}, n^{(2)}$ is given by
$$
n^{(1)}=\left(1,1,1/c_3 \right)\in A^+,\quad n^{(2)}=Wn^{(1)}=\left(c_1,1,1 \right)\in A^+,
$$
which ends the proof.
$\hfill\blacksquare$
\vspace{5mm}

\noindent \textbf{Proof of Theorem \ref{th:sing-gen-odd} (b): Configurations without Type 2 singularities.}\\
\noindent Assume $m=3$ and let $c_1, c_2, c_3\ne 0$ be specified weights. By Remark \ref{rem:sign-pos}, each stable configuration of $\mathcal{S}$ with signed weights $c_1, c_2, c_3$ corresponds to another stable configuration of $\mathcal{S}$ with positive weights $|c_1|$, $|c_2|$, $|c_3|$ and the other way around. Hence, without loss of generality we will prove our statement for $c_1,c_2,c_3>0$.

We will use our knowledge of existence of stable configurations for every star transform with uniform weights (marked by superscript $^{(2)}$ in the proof below) to show existence of stable configurations for star transforms with positive weights (marked by superscript $^{(1)}$ in the proof below).

By Lemma \ref{lem:plane-trans}  there exist  SSAV planes $T_1,T_2$ such that $T_2=W^{-1}T_1$ and $T_i\cap e_2^{-1}(0)=\{0\}$ for $i=1,2$. Let $a^{(1)},b^{(1)}$ and $a^{(2)},b^{(2)}$ be some star aperture vector pairs spanning correspondingly $T_1$ and $T_2$.

To prove the theorem, it is enough to show that for ray directions $\gamma_1^{(1)}, \gamma_2^{(1)}, \gamma_3^{(1)}$ corresponding to the aperture vectors $a^{(1)},b^{(1)}$  the following relation holds
$$
P_2^{(1)}(\psi)\doteq c_1\langle \psi, \gamma_2^{(1)} \rangle\langle \psi, \gamma_3^{(1)} \rangle+c_2\langle \psi, \gamma_1^{(1)} \rangle\langle \psi, \gamma_3^{(1)} \rangle+c_3\langle \psi, \gamma_1^{(1)} \rangle\langle \psi, \gamma_2^{(1)} \rangle\ne0
$$
for all $\psi\in \mathbb{S}^1$.


Due to the properties of selected planes and corresponding aperture vectors
\begin{equation*}
P_2^{(2)}(\psi)\doteq e_2\left(ra^{(2)}+sb^{(2)}\right)\ne0
\mbox{ for all } \psi=(r,s)\in \mathbb{S}^1.
\end{equation*}
At the same time for any $\psi=(r,s)\in \mathbb{S}^1$ there exists $\hat\psi=(\hat{r},\hat{s})\in \mathbb{S}^1$ such that
\begin{equation*}
P_2^{(1)}(\psi)=c\;e_2\left(W^{-1}(ra^{(1)}+sb^{(1)})\right)=k\;e_2\left(\hat{r}a^{(2)}+\hat{s}b^{(2)}\right)=k\;P^{(2)}_2(\hat\psi)\ne0,
\end{equation*}
where $c=c_1c_2c_3$ and $k$ is some positive constant. In the second equality above we use the fact that $W^{-1}(ra^{(1)}+sb^{(1)})\in T_2$ and homogeneity of $e_2$.
$\hfill\blacksquare$

\vspace{5mm}

The generalization of Theorem \ref{th:sing-gen-odd} (b) to the case of $m=2k+1>3$ is technically more complicated and we plan to address it in a future publication. So we finish this sections with the following.

\begin{conjecture}
 Let $m=2k+1$. For any set of specified weights $c_1, \ldots, c_m$, there exist $\gamma_1,\ldots,\gamma_m$ such that $\mathcal{S}$ does not contain singular directions of Type 2. In other words, for any set of weights there are configurations of $\mathcal{S}$ with a stable inversion.
\end{conjecture}

\section{Zeros of $e_{m-1}(y_1,\ldots, y_m)$ for Odd $m$.}\label{sec:conj}

In paper \cite{Conflitti} published in 2006, the author formulated the following conjecture:
\begin{conjecture}
 If $r$ is even then $e^{-1}_r(0)$ contains no real vector subspace of dimension $r$.
\end{conjecture}
Furthermore, it is stated there that one of the extreme cases is ``the case $e_{m-1} (y_1,...,y_m),\, m \equiv 1\; (\operatorname{mod} 2)$, which becomes a task quite hard to tackle''.

\vspace{3mm}

\noindent Our proof of Theorem \ref{th:reg} includes a proof of the aforementioned extreme case.  Namely,

\begin{theorem}\label{th-zeros}
Let $m=2k+1$ for some $k\in\mathbb{N}$. Then $e_{m-1}^{-1}(0)$ contains no real vector subspace of dimension $m-1$.
\end{theorem}

Indeed, let $a,b\in\mathbb{R}^m$ be the (linearly independent) aperture vectors of the regular star with $m=2k+1$ rays. It was shown that
$e_{m-1}(ra+sb)=P_2(\psi)\ne 0$ for any $\psi=(r,s)\in \mathbb{S}^1$. The set $C\doteq\left\{ra+sb:\,(r,s)\in \mathbb{S}^1\right\}$ represents a closed contour around the origin in the 2-dimensional subspace $T\subset\mathbb{R}^m$ spanned by the aperture vectors $a$ and $b$. Due to homogeneity of $e_{m-1}(y)$, the fact that it has no zeros on $C$, implies that $e^{-1}_{m-1}(0)\bigcap T=\{0\}$. Since $\dim(T)=2$, the zero set $e^{-1}_{m-1}(0)$ can not contain a real vector subspace of dimension $m-1$. $\hfill\blacksquare$


\section{The Star Transform in Imaging} \label{sec:applications}
\begin{figure}[h]
\begin{center}
\includegraphics[height=1.8in]{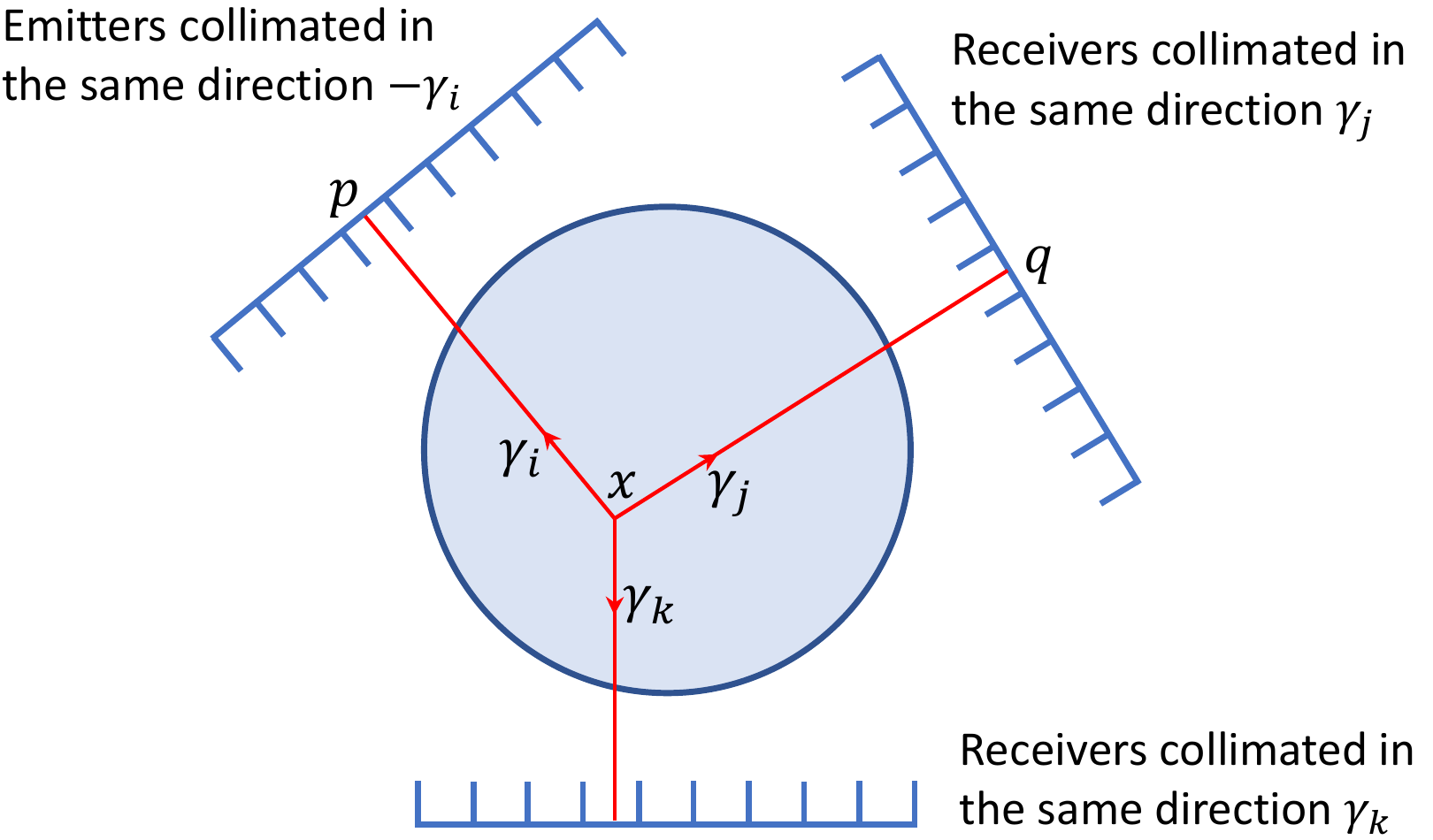}
\caption{A simple setup of single-scattering tomography.} \label{fig9}
\end{center}
\end{figure}

A known application of the star transform is related to imaging of a heterogeneous medium using single-scattered radiation \cite{ZSM-star-14}. To describe a simple model of such procedure, consider the following setup (see Figure \ref{fig9}).

Let $p$ and $q$ be some points outside the imaged body. A source of radiation located at point $p$ emits a beam of particles into the body along the direction $-\gamma_i$.  A receiver placed at point $q$ collimated along the direction $\gamma_j$ measures the intensity of scattered radiation. Ignoring particles that scatter more than once and have very small energy at $q$, each collimated emitter-receiver pair $(p, q)$ uniquely identifies the scattering location $x$ inside the body. That point $x$ corresponds to the vertex of the V-line with rays along $\gamma_i, \gamma_j$ passing respectively through $p, q$. The measured data $\phi_{ij}(x)$ can then be expressed as
    \begin{equation}\label{scat-absorp}
     \phi_{ij}(x) = \mathcal{X}_i f(x)+ k_{ij}\mathcal{X}_j f(x) + \eta{(x)},
    \end{equation}
where $f(x)$ and $\eta(x)$ represent correspondingly the attenuation and scattering coefficients at $x$, while $k_{ij}$ is a known constant. Relation (\ref{scat-absorp}) can be  derived using single-scattering approximation to the radiative transport equation  (see \cite{Florescu2, Kats-Kryl-15, ZSM-star-14}).

In simple models of single scattering tomography it is customary to take $k_{ij}=1$, assuming that the medium has a unique (space dependent) attenuation coefficient. However, it is well known that the attenuation is energy dependent. Given the drop of energy of the particles after scattering, it is natural to consider different $f$'s before and after scattering. For example, the primary type of scattering in X-rays is Compton scattering, where the energy loss depends on the scattering angle. If one collects data corresponding to two fixed directions $\gamma_i, \gamma_j$, then (under certain simplifying assumptions) it is reasonable to consider the attenuation coefficients at different energy levels (before and after the single scattering event) as $f$  and $k_{ij} f$ (e.g. see \cite{Kats-Kryl-15}).

Let us now return to formula (\ref{scat-absorp}). By varying the location of points $p, q$ and using finitely many fixed directions for $\gamma$'s (see Figure \ref{fig9}), one can generate a two-dimensional family of measured data corresponding to all possible locations of $x$ inside the body. If $\eta(x)$ is known, then the recovery of $f(x)$ boils down to inversion of a weighted V-line transform. In this case just two fixed directions $\gamma_1, \gamma_2$ will suffice.

However, even in the simple model using $k_{ij}=1$, if $\eta(x)$ is not known, the recovery of both attenuation and  scattering coefficients from the same data set requires the use of the star transform with at least three distinct ray directions. The key idea here is to consider various linear combinations of $\phi_{ij}(x)$, which will eliminate $\eta$ and produce a weighted star transform of $f$. For example, in a setup using three fixed directions $\gamma_1, \gamma_2, \gamma_3$ one can generate star transform data as follows:
$$
\phi_{12}(x)-\phi_{23}(x)=\mathcal{X}_1 f(x) + (k_{12}-1)\mathcal{X}_2 f(x)+k_{23} \mathcal{X}_3 f(x).
$$
An inversion of the weighted star transform produces $f(x)$, which then can be used to recover $\eta(x)$ from (\ref{scat-absorp}).

The choice of a linear combination of $\phi_{ij}(x)$ eliminating $\eta(x)$ is not unique, even when $k_{ij}=1$ for all $i,j$. For example, let us assume that the existing hardware allows measurements along $m$ directions $\gamma_1, \ldots, \gamma_m$.
To get an appropriate linear combination of data at each vertex $x$ and cancel the contribution from the $\eta$ term, one can choose a symmetric matrix $\omega$ with zeros on the main diagonal and satisfying $\sum_{i,j} \omega_{ij} = 0$. Writing this linear combination, we get
    $$
     \sum_{i=1}^m \sum_{j=1}^m \omega_{ij} \phi_{ij}(x) = \sum_{i=1}^m \left(\sum_{j=1}^m \omega_{ij}\right) \mathcal{X}_i f(x) = \sum_{i=1}^m c_i \mathcal{X}_i f(x),
    $$
    where $$c_i=\sum_{j=1}^m \omega_{ij}.$$
After applying the inversion of the star transform and finding $f$, we can recover the scattering term using (\ref{scat-absorp}).

It is easy to notice that when $m>3$, for any predefined set of weights $c_i\ne0$, $i=1,\ldots,m$ there are infinitely many choices of $\omega$ satisfying the described requirements. In the case of $m=3$, not all choices of $c_1, c_2, c_3$ work, but there are still infinitely many admissible options.

We refer the interested reader to \cite{Kats-Kryl-15, ZSM-star-14} for more details on physical aspects and imaging settings of single-scattering optical and X-ray tomography.

\begin{remark}
     Another potential imaging application of V-line transforms is emission tomography, where the goal is recovery of the distribution $f(x)$ of a radioactive material inside an object. Here the measured data corresponding to scattered particles can be represented by the weighted V-line transforms of $f(x)$, where the non-constant weight will depend on a spatially varying attenuation $\mu(x)$. To the best of our knowledge, no inversion formula is known for such V-line transforms with non-constant weights.
\end{remark}


\section{Numerical Simulations}\label{sec:numerical}

\begin{figure}[H]
	\begin{center}
		\begin{subfigure}{0.48\textwidth}
			\includegraphics[height=39mm,keepaspectratio]{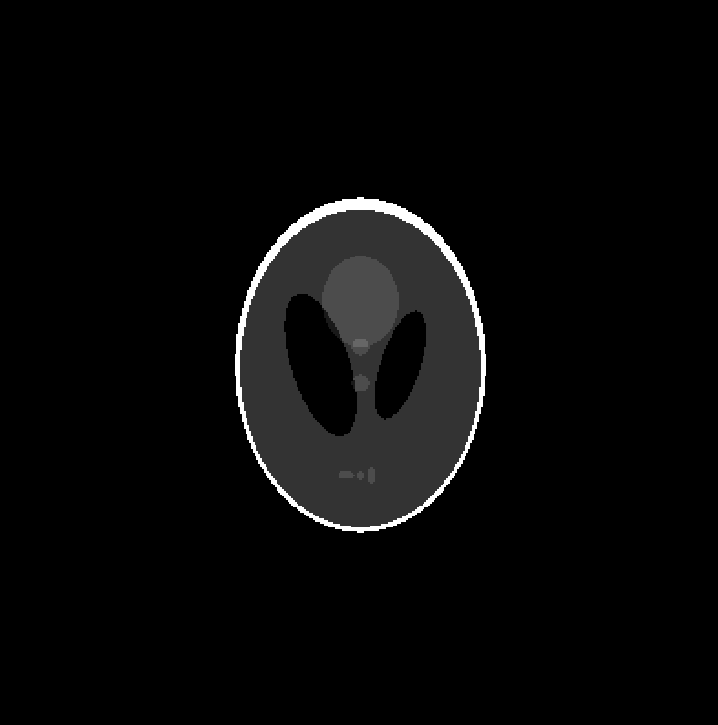}
		\caption{Shepp-Logan phantom\\ used in all reconstructions }	
		\end{subfigure}	
		\begin{subfigure}{0.48\textwidth}
            \includegraphics[height=39mm,keepaspectratio]{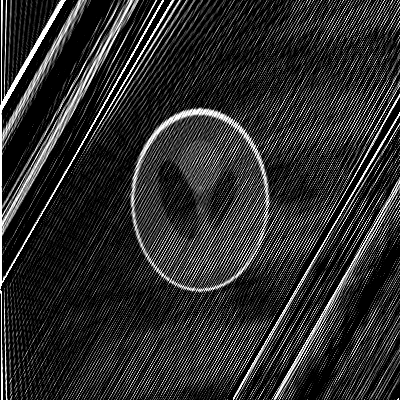}	
        \caption{Reconstruction from $\mathcal{S}$ with\\ $\gamma_1=(1,0), \gamma_2=(\cos\frac{2\pi}{3},\sin\frac{2\pi}{3})$}
        \end{subfigure}
	\caption{}
	\label{fig:stfig11}
	\end{center}		
\end{figure}

In this section we present some examples of numerical reconstructions of the standard Shepp-Logan phantom (see Figure \ref{fig:stfig11}a) from its star transforms of various configurations. The reconstruction algorithm is based on Theorem \ref{Th:StarInversion} and the filtered backprojection algorithm of the Radon transform. The resolution of all images is 400x400 pixels. The weights $c_i=1$ for all $i$.

In all simulations below we have generated the star transform $\mathcal{S}f$ of $f$ depicted in Figure \ref{fig:stfig11}a by numerically computing the divergent beam transforms $\mathcal{X}_{\gamma_j} f$ along given directions $\gamma_j$ and adding them up. Since all $\mathcal{X}_{\gamma_j} f$'s and $\mathcal{S}f$ have unbounded support (even for compactly supported $f$), one has to truncate the data. In other words, numerically $\mathcal{S}f$ is represented by a matrix $A$, where each entry $a_{i,j}$ corresponds to $\mathcal{S}f(x_{i,j})$ with vertices $x_{i,j}$ uniformly sampled inside the square $[-1,1]\times[-1,1]$, while $\mathcal{S}f$ has unbounded support. Such truncation creates numerical errors when computing the Radon transform of $\mathcal{R}(\mathcal{S}f)(\psi,{t})$ along lines that pass through truncated ``tail'' of $\mathcal{S}f$. In the reconstructed images these errors appear in the form of artifacts at the edges of the unit square. Moreover, in unstable geometric configurations those errors get amplified along lines $l(\psi,t)$ with normal unit vector $\psi$ corresponding to singular directions of Type 2 for $\mathcal{S}$, due to the presence of $q(\psi)$ in the inversion formula (\ref{eq:Rad-Star}).

It is easy to notice that in the setup corresponding to an even number of rays  (see Figure \ref{fig:stfig11}b) the reconstruction has severe artifacts propagating along lines with normals in the direction of Type 2 singularities. Namely, the directions of Type 2 singularities are $\psi_1=(\cos\frac{\pi}{6}, -\sin\frac{\pi}{6})$ and $\psi_2=(\cos\frac{5\pi}{6}, \sin\frac{5\pi}{6})$, and the severe artifacts are pronounced along lines parallel to the vector $(\cos\frac{\pi}{3}, \sin\frac{\pi}{3})$. Similar issues can be observed in  Figure  \ref{fig:stfig51} corresponding to the configuration with odd number of rays and Type 2 singularities described in the proof of Part 2a of Theorem \ref{th:singularities}.

\begin{figure}[H]
	\begin{center}
		\begin{subfigure}{0.48\textwidth}
		\includegraphics[height=39mm,keepaspectratio]{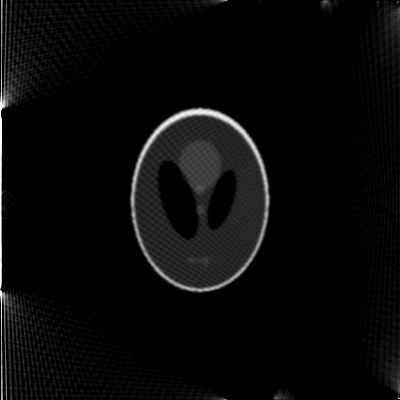}
			\caption{Reconstruction from regular $\mathcal{S}$\\ with $\gamma_1=(1,0), \gamma_2=(\cos\frac{2\pi}{3},\sin\frac{2\pi}{3}),\\\gamma_3=(\cos\frac{4\pi}{3},\sin\frac{4\pi}{3})$ }
			\label{fig:stfig31}
		\end{subfigure}	
	    \begin{subfigure}{0.49\textwidth}	
	    \includegraphics[height=39mm,keepaspectratio]{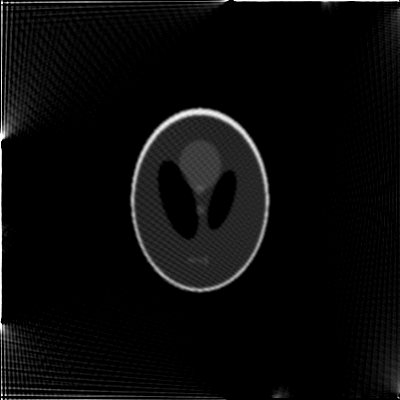}
			\caption{Reconstruction from $\mathcal{S}$ with\\  $\gamma_1=(\cos\frac{\pi}{20},\sin\frac{\pi}{20}), \gamma_2=(\cos\frac{2\pi}{3},\sin\frac{2\pi}{3}),\\ \gamma_3=(\cos\frac{4\pi}{3},\sin\frac{4\pi}{3})$}
			\label{fig:stfig41}
	    \end{subfigure}	
	\caption{}
	\end{center}		
\end{figure}

\begin{figure}[H]
	\begin{center}
		\begin{subfigure}{0.48\textwidth}
		\includegraphics[height=39mm,keepaspectratio]{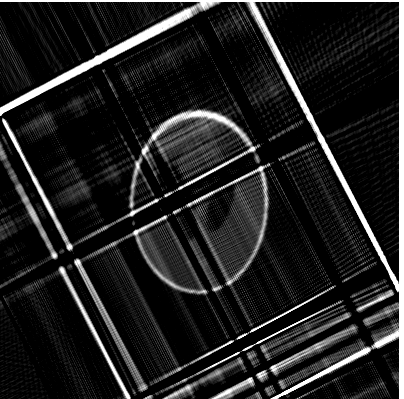}
			\caption{Reconstruction from $\mathcal{S}$ with\\ $\gamma_1=(1,0), \gamma_2=(0,1),\\ \gamma_3=(\cos\frac{3\pi}{4},\sin\frac{3\pi}{4})$\\  }
			\label{fig:stfig51}
		\end{subfigure}	
	    \begin{subfigure}{0.49\textwidth}	
		\includegraphics[height=39mm,keepaspectratio]{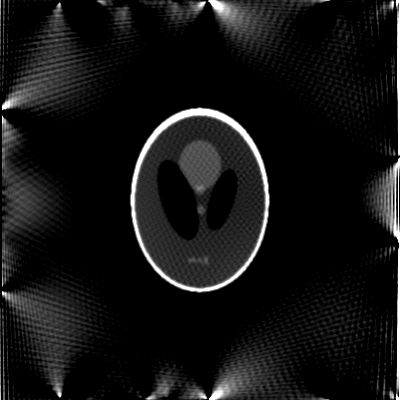}
			\caption{Reconstruction from regular $\mathcal{S}$ with\\ $\gamma_1=(1,0)$, $\gamma_2=(\cos\frac{2\pi}{5},\sin\frac{2\pi}{5}),$\\
			$\gamma_3=(\cos\frac{4\pi}{5},\sin\frac{4\pi}{5}),$ $\gamma_4=(\cos\frac{6\pi}{5},\sin\frac{6\pi}{5}),$\\ $\gamma_5=(\cos\frac{8\pi}{5},\sin\frac{8\pi}{5})$ }
			\label{fig:stfig61}
       \end{subfigure}	
	\caption{}
	\end{center}		
\end{figure}

The setups with an odd number of rays that do not have singular directions of Type 2 are of much better quality and do not have such severe artifacts (see Figures \ref{fig:stfig31}, \ref{fig:stfig41} and \ref{fig:stfig61}).The milder artifacts close to the boundary of the unit square are almost entirely outside of the support of the image function (i.e. outside of the unit disc) and could have been cleared after the reconstruction, but we kept them for completeness of the experimental results.

\vspace{-7mm}
\vspace{-7mm}

\section{Additional Remarks}\label{sec:remarks}
\begin{enumerate}
    \item The statement of a special case of Theorem \ref{Th:StarInversion} was published without proof in the abstract \cite{Amb-Latifi2}.

    \item We did not aim to formulate the most general conditions under which Theorem \ref{Th:StarInversion} holds. The requirement of $f\in C_c(\mathbb{R}^2)$ is sufficient, but not necessary. The proof will hold under weaker assumptions, e.g. when $f$ decays sufficiently fast at infinity.

    \item

    In the case of $m=1$, Theorem \ref{Th:StarInversion} establishes a relation between the divergent beam transform  and the (ordinary)  Radon transform. A different relation between these transforms was obtained in \cite{hamaker_et_al}, where the divergent beam transform is integrated over all possible angles in $S^{n-1}$ to generate the integral of the function over the whole space $\mathbb{R}^n$. The latter can also be obtained by integrating the Radon transform over all hyperplanes normal to a given direction.

    Similarly, in the case of $m=2$, our Theorem \ref{Th:StarInversion} establishes a relation between the V-line transform and the Radon transform. A different relation between these transforms was obtained in \cite{Terz} using an overdetermined setup of the V-line transform. It is similar in the spirit to the formula for the divergent beam transform in \cite{hamaker_et_al}, as it uses all possible angles for the branch directions of the V-line transform.

    Our formulas derived in Theorem \ref{Th:StarInversion} are essentially different from those in \cite{hamaker_et_al} and \cite{Terz}, as we do not require overdetermined data and use only a finite number of branch directions.

\end{enumerate}

\section{Acknowledgements}\label{sec:acknowledge}

We would like to thank the anonymous reviewers for several insightful observations that allowed us to shorten some of the proofs and considerably improve the manuscript.

This work was partially funded by NSF grant DMS 1616564 and Simons Foundation grant 360357. The authors are thankful to Alessandro Conflitti for updates about the current status of his conjecture and useful references.

\end{document}